\documentclass[conference,a4paper]{IEEEtran}
\usepackage{verbatim}
\usepackage{cite}
\usepackage{amsmath}
\usepackage{amsfonts}
\usepackage{url}
\usepackage{graphicx}
\usepackage{subfigure}
\usepackage{caption}
 \usepackage{epstopdf}
\usepackage{multirow}
\usepackage{color,xcolor}
\usepackage{CJK}
\usepackage{comment}
\usepackage{booktabs}
\usepackage{subfigure}

\AtBeginDocument{%
  \providecommand\BibTeX{{%
    \normalfont B\kern-0.5em{\scshape i\kern-0.25em b}\kern-0.8em\TeX}}}

\IEEEoverridecommandlockouts
\IEEEpubid{\makebox[\columnwidth]{978-1-6654-7592-1/22/\$31.00~\copyright~2022 IEEE \hfill} \hspace{\columnsep}\makebox[\columnwidth]{ }}

\begin{document}
%
\title{Reduced Reference Quality Assessment for Point Cloud Compression}

\author{\IEEEauthorblockN{Yipeng Liu}
\IEEEauthorblockA{\textit{Cooperative MediaNet Innovation Center} \\
\textit{Shanghai Jiao Tong University} \\
Shanghai, China \\
liuyipeng@sjtu.edu.cn}
\and
\IEEEauthorblockN{Qi Yang}
\IEEEauthorblockA{\textit{Cooperative MediaNet Innovation Center} \\
\textit{Shanghai Jiao Tong University} \\
Shanghai, China \\
yang\_littleqi@sjtu.edu.cn}
\and
\IEEEauthorblockN{Yiling Xu}
\IEEEauthorblockA{\textit{Cooperative MediaNet Innovation Center} \\
\textit{Shanghai Jiao Tong University} \\
Shanghai, China \\
yl.xu@sjtu.edu.cn}

}

\maketitle

\begin{abstract}

\par In this paper, we propose a reduced reference (RR) point cloud quality assessment (PCQA) model named R-PCQA to quantify the distortions introduced by the lossy compression. Specifically, we use the attribute and geometry quantization steps of different compression methods (i.e., V-PCC, G-PCC and AVS) to infer the point cloud quality, assuming that the point clouds have no other distortions before compression. First, we analyze the compression distortion of point clouds under separate attribute compression and geometry compression to avoid their mutual masking, for which we consider 5 point clouds as references to generate a compression dataset (PCCQA) containing independent attribute compression and geometry compression samples. Then, we develop the proposed R-PCQA via fitting the relationship between the quantization steps and the perceptual quality. We evaluate the performance of R-PCQA on both the established dataset and another independent dataset. The results demonstrate that the proposed R-PCQA can exhibit reliable performance and high generalization ability.


\end{abstract}


%
\IEEEpeerreviewmaketitle

\section{Introduction}
\label{sec:introduction}

\par Recently, point cloud has emerged as a promising representation format in prevalent 3D applications (e.g., autonomous driving~\cite{Li2021ADrive} and augmented reality~\cite{Lim2020AR}), for which the point cloud compression (PCC) is of great interest for providing efficient service  in practices. Currently, the Moving Picture Experts Group (MPEG) has applied the separable measurements of geometry and attribute distortion in the course of lossy PCC. For the geometry distortion, MPEG proposes to use  the point-to-point (p2point)~\cite{cignoni1998metro}, or point-to-plane (p2plane)~\cite{Tian2017Evaluation} to quantify the spatial perturbation; while for the attribute part, the PSNRyuv is proposed to measure the differences between corresponding color channels. Besides these metrics which have already been applied in MPEG PCC standardization, some other metrics which consider more human visual characteristics and present better performance on public PCQA databases are also developed, such as ~\cite{torlig2018novel,alexiou2018pointt,Yang2020TMM3DTO2D,javaheri2021JPC, meynet2019pcmsdm,viol2020acolor,alexiou2020TowardsStructural,meynet2020pcmd,yang2020graphsim,Zhang2021MSGraphSIM,javaheri2021PTD}. However, they are full reference (FR) metrics which require both the reference and distorted samples and have high computational complexity for real-time quality prediction.

\par In many practical cases, e.g., transmission, the timely feedback is expected, and only the compressed samples and the meta data are available, in which the reduced reference (RR) PCQA metrics are indispensable. Only a few researches explore the RR methods for PCQA. \cite{Irene2020RR2} uses the statistical information (e.g., the luminance histogram) as the substitute for the complete samples, but still requires the backend processing. \cite{Liu2021RR} applies the quantization parameters in V-PCC to estimate the quality of compressed samples and guide rate control, but other prevalent compression strategies (e.g., G-PCC) are ignored. Therefore, in this paper, we propose a general RR PCQA model for compression distortions named R-PCQA which only takes the attribute and geometry quantization steps of compression schemes (including  V-PCC~\cite{VPCC}, G-PCC~\cite{GPCC} and AVS~\cite{AVS}, V-PCC and G-PCC are provided by MPEG while AVS is recommended by China Audio-Video Coding Standard) as variables, since the quality of compressed point clouds is highly related to the compression parameters. To fully study the relationship between the compression parameters and perceptual quality, we first establish a complete subjective database for PCC, named PCC quality assessment (PCCQA) database.

\par The reason why we establish PCCQA while many PCQA datasets~\cite{Alexiou2020PointXR,Yang2020TMM3DTO2D,Javaheri2019IRPC,Liu2022LSPCQA,Su2019WPC} have been proposed is that current databases only consider the superimposed compression distortion, i.e. lossy-geometry (G)-lossy-attribute (A) compression, which is recommended in the Common Test Conditions (CTC) \cite{VPCC-CTC,GPCC-CTC,AVS-CTC}. The separate compression strategies, i.e. lossless-G-lossy-A and lossy-G-lossless-A compression, which are not included in the CTC are often ignored. Considering the mutual masking between geometry and attribute distortions \cite{sub6-javaheri2020point}, they are useful for exploring the relationship between the perceptual quality and compression parameters. In PCCQA, the reference point clouds are compressed by V-PCC, G-PCC and AVS under lossless-G-lossy-A condition, lossy-G-lossless-A condition, and lossy-G-lossy-A condition respectively.

 To model the relationship between the perceptual quality and the compression parameters, we first convert all the compression parameters to the quantization steps. Then, we model the relationship between the perceptual quality and the attribute/geometry quantization steps respectively  via using the least square fitting. Finally, the proposed R-PCQA combines the attribute compression model and geometry compression model to predict the final quality scores.

\par The rest of this paper is organized as follows: section \ref{sec:dataset} introduces the established PCCQA dataset; section \ref{sec:model} presents the proposed R-PCQA; section \ref{sec:experiment} illustrates the experiment results; the conclusion is summarized in section \ref{sec:conclusion}.

\section{PCCQA database}
\label{sec:dataset}

\par To better explore the relationship between the perceptual quality and the attribute/geometry compression parameters, we first establish a database called PCCQA under several compression conditions.

\par Five reference point clouds are selected from MPEG and AVS point cloud datasets. These reference point clouds are ensured to have no holes and other distortions under 1080P presentation with size-2 primitives. The reference point clouds are then distorted with 3 compression algorithms, i.e. V-PCC~\cite{VPCC}, G-PCC~\cite{GPCC} and AVS~\cite{AVS}. Each compression is conducted under 3 conditions, i.e. lossless-G-lossy-A, lossy-G-lossless-A, and lossy-G-lossy-A. The compression parameters are shown in Table \ref{alldistortions}. In total, 225 compressed point clouds are generated.


\begin{table}[htbp]
\tiny
\caption{Compression parameters used for distorted point cloud generation.}
\begin{tabular}{lllll}
\hline
& Conditions & Parameters &  &  \\ \hline
             & \begin{tabular}[c]{@{}l@{}}GPCC lossy-G-lossless-A\\      \\ VPCC lossy-G-lossless-A\\ \\ AVS lossy-G-lossless-A \end{tabular} & \begin{tabular}[c]{@{}l@{}}(positionQuantizationScale)\\ 0.75 0.5 0.25 0.125 0.0625\\ (geomQP) \\ 22 32 37 42 51\\ (geom\_quant\_step)\\ 2 4 8 12 16\end{tabular}                &  &  \\\hline
                    & \begin{tabular}[c]{@{}l@{}}GPCC lossless-G-lossy-A\\      \\ VPCC lossless-G-lossy-A\\ \\ AVS lossless-G-lossy-A\end{tabular}             & \begin{tabular}[c]{@{}l@{}}(qp) \\ 35 39 43 47 51\\ (textureQP) \\ 32 37 42 47 51\\ (attr\_quant\_param)\\ 24 32 40 44 48\end{tabular}                                                                              &  &  \\ \hline
 & \begin{tabular}[c]{@{}l@{}}GPCC lossy-G-lossy-A\\      \\ VPCC lossy-G-lossy-A\\ \\ AVS lossy-G-lossy-A\end{tabular}                   & \begin{tabular}[c]{@{}l@{}}(positionQuantizationScale, qp)   \\ 0.75,35 0.5,39 0.25,43 0.125,47 0.0625,51\\ (geomQP, textureQP) \\ 24,32 28,37 32,42 36,47 40,51\\ (geom\_quant\_step, attr\_quant\_param)\\ 2,24 4,32 8,40 12,44 16,48\end{tabular} &  &  \\ \hline
\end{tabular}
\label{alldistortions}%
\end{table}

\par To annotate the compressed point clouds, a subjective experiment is organized to collect the Mean Opinion Scores (MOS). We adopt the double stimulus method since it can obtain more stable results for minor impairments. The experiment process and environment setting strictly follow the ITU-R Recommendation BT. 500~\cite{BT500}. Such a method for collecting subjective MOS is also adopted in other researches, such as~\cite{Wu20206Dof,Su2019PQA,Liu2022LSPCQA,Yang2021ITPCQA}.


\section{Proposed Quality Assessment Model R-PCQA}
\label{sec:model}

\subsection{Unifying the Compression Parameters}

\par The compression parameters with different meanings are used in V-PCC, G-PCC and AVS, but the used compression parameters can all be converted to the quantization steps. Thus, to better explore the relationship between the perceptual quality and the quantization, we first convert these compression parameters to quantization steps, denoted as $Qs$, before proposing the R-PCQA.

\par In V-PCC, the parameters \textbf{textureQP} and \textbf{geomQP}, denoted as $QP$, are used to control the attribute compression and geometry compression respectively, which apply
\begin{align}\label{v-pcc-pq}
Qs = round({2^{\frac{{QP - 4}}{6}}}),
\end{align}
where $round(\cdot)$ means converting a number to the nearest integer.

\par In attribute compression of G-PCC, the compression parameter \textbf{qp} has the same meaning as $QP$ in V-PCC, following the same conversion formula in Eq. \eqref{v-pcc-pq}. The parameter \textbf{positionQuantizationScale}, denoted as $S$, is used to control the geometry quantization, which can be converted to $Qs$ by
\begin{align}\label{gpcc-qp}
{Qs = }\frac{1}{{S}}.
\end{align}
\par In AVS, the parameter \textbf{attr\_quant\_param}, denoted as $QP_a$, is used to control the attribute quantization, which can use the following formulation to convert it to $Qs$
\begin{align}
Qs = {2^{\frac{{QP_a}}{8}}}.
\end{align}
\par For the geometry compression of AVS, the  parameter \textbf{geom\_quant\_step} shares the same meaning with the quantization step $Qs$.

\begin{figure}[t]
\centering
  \subfigure[]{\includegraphics[width=0.7\linewidth]{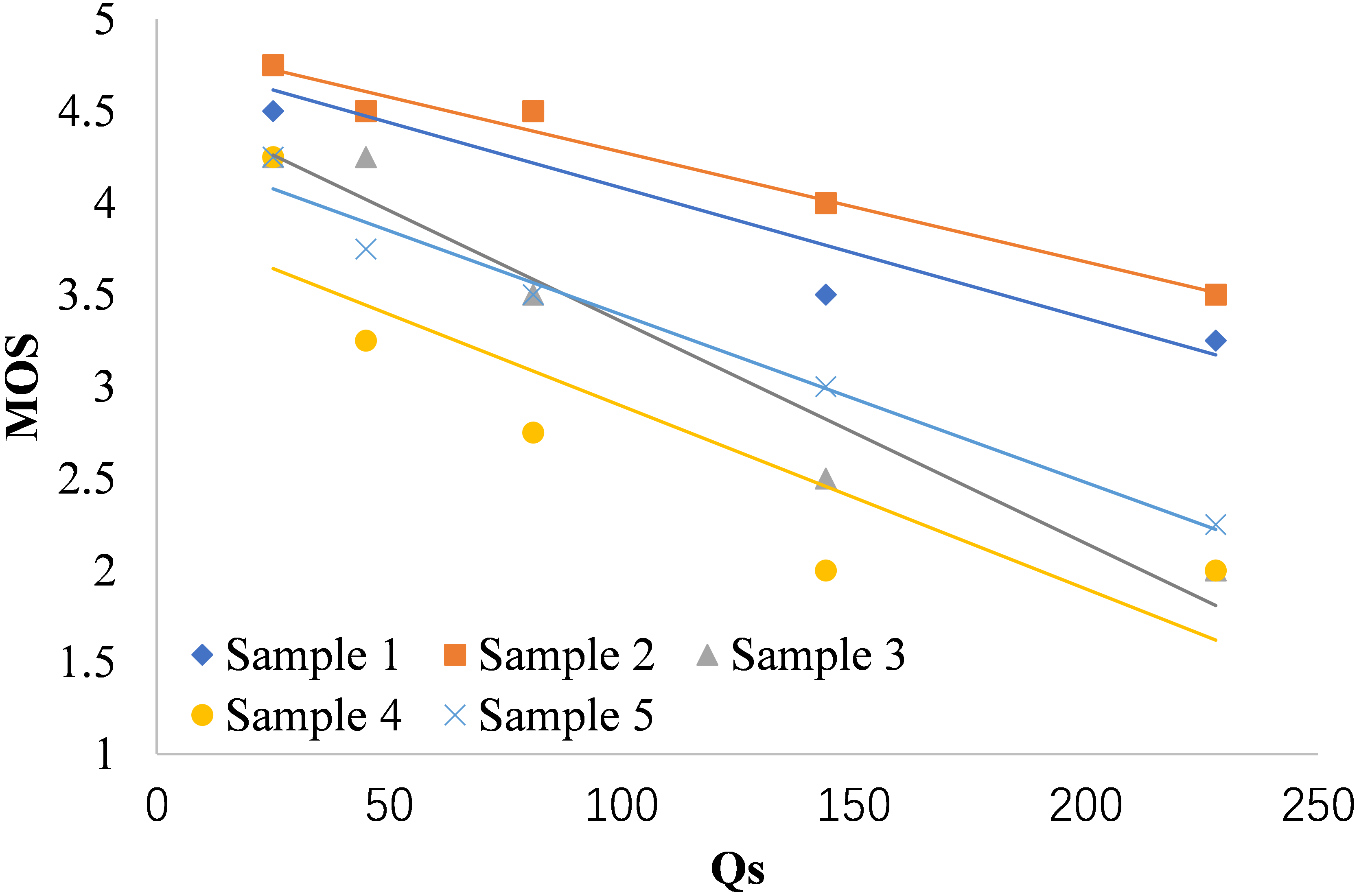}\label{sfig:predictionErrorPartI}}
  \subfigure[]{\includegraphics[width=0.7\linewidth]{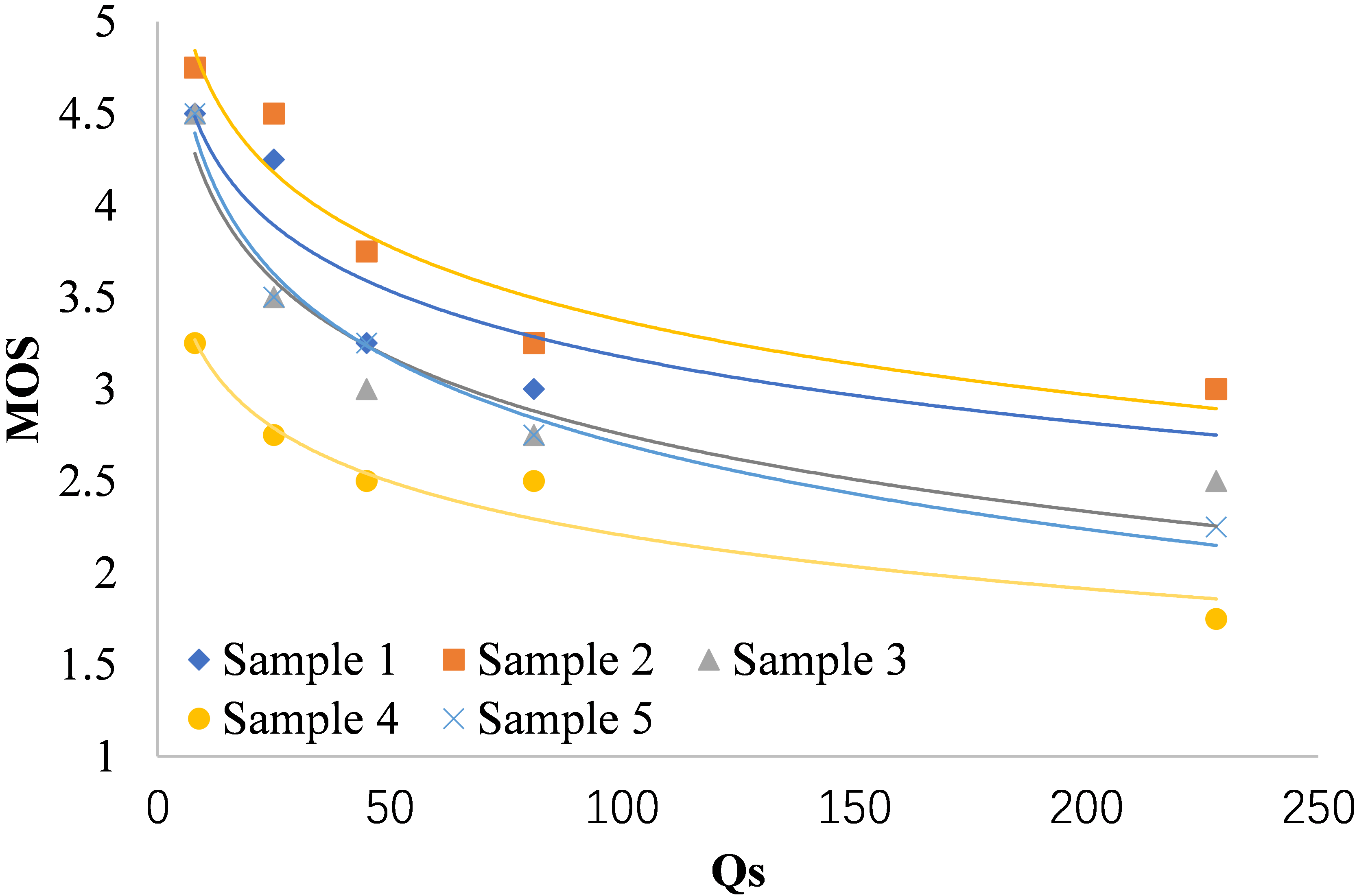}\label{sfig:predictionErrorPartII}}
  \centering
\caption{{\bf Variation of Qs as function of MOS for different samples.} (a) under V-PCC lossless-G-lossy-A condition; (b) under V-PCC lossy-G-lossless-A condition.}
  \label{Qs2MOS}
\end{figure}

\subsection{Overall Quality Model}
\label{overallmodel}


We use the average MOS in PCCQA to fit the mathematical model for quality prediction. The relationships between $MOS$ and $Qs$ under V-PCC lossless-G-lossy-A condition and V-PCC lossy-G-lossless-A condition are shown in Fig.~\ref{Qs2MOS}. We can see that under the same compression condition, different samples share the fitting model with basically the same shape but are added to different additive factors. Therefore, we assume $MOS$ and $Qs$ satisfy the following relationship  under a certain compression condition:
\begin{align}
\label{formu1}
MOS = F(Qs) + c(pc),
\end{align}
where $F$ denotes the fitting function which is related to the quantization step $Qs$. $c$ denotes the additive factor which is related to the intrinsic characteristics of the point cloud $pc$.

\par On the whole, different samples share the same relationship model under a certain compression condition, but they are added to an additive sample factor. To deal with the additive sample factor, we use $Qs$ and  average $MOS$ which is denoted as $\overline{MOS}$ to build up the relationship model for each compression condition:
\begin{align}
\label{formu1}
MOS_f = \overline{MOS} = F(Qs) + \overline{c(pc)},
\end{align}
where $MOS_f$ is the final predicted quality score and $\overline{c}$ denotes the average value of additive factors.

\subsection{Modeling the Attribute Compression}

\par
\begin{figure*}[htbp]
\centering
  \subfigure[]{\includegraphics[width=0.25\linewidth]{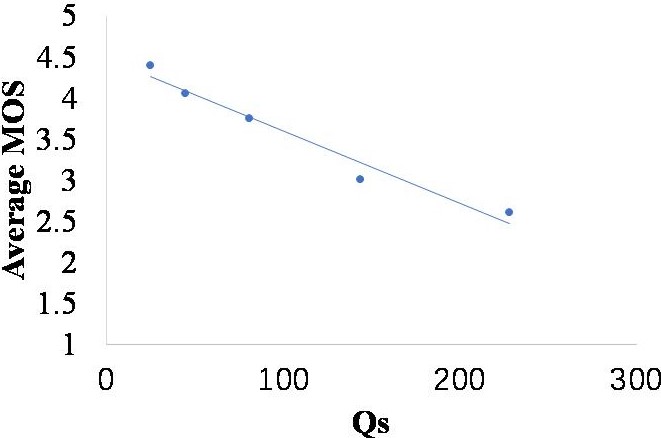}}
  \subfigure[]{\includegraphics[width=0.25\linewidth]{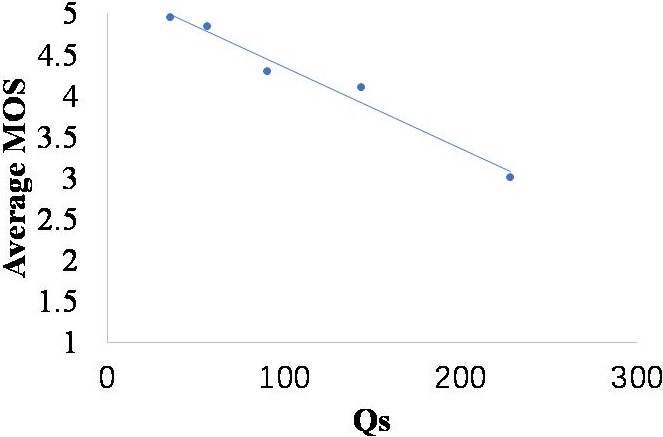}}
  \subfigure[]{\includegraphics[width=0.25\linewidth]{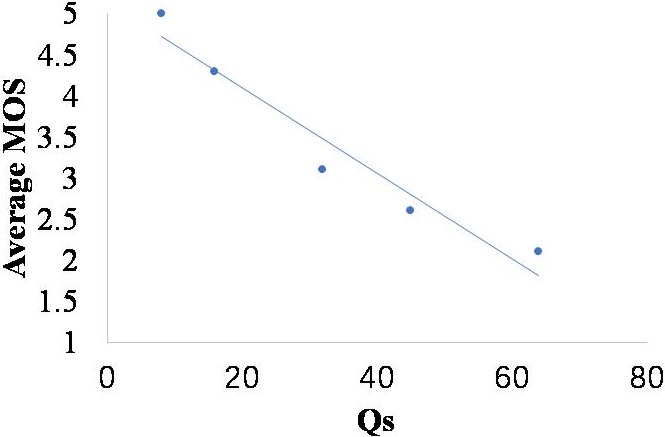}}
  \subfigure[]{\includegraphics[width=0.25\linewidth]{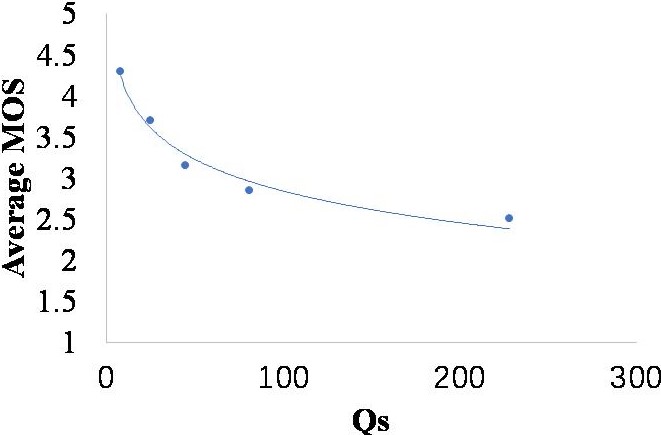}}
  \subfigure[]{\includegraphics[width=0.25\linewidth]{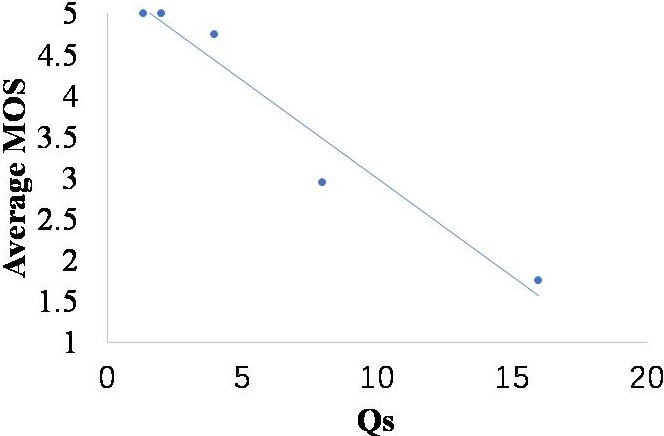}}
  \subfigure[]{\includegraphics[width=0.25\linewidth]{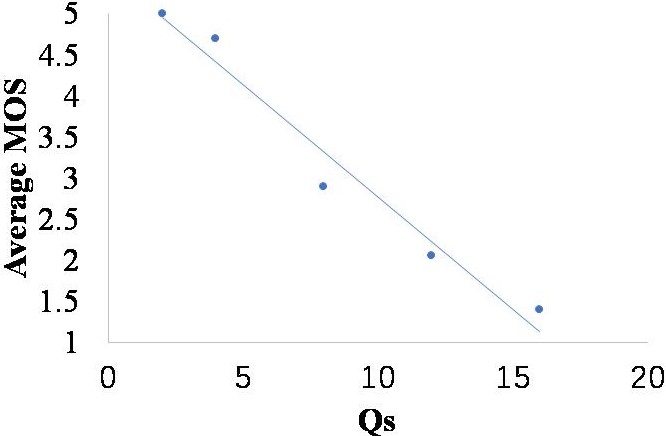}}
  \centering
\caption{{\bf Variation of Qs as function of average MOS for each compression condition.} The top row is under lossless-G-lossy-A, and the bottom row is under lossy-G-lossless-A. (a) (d) is for V-PCC, (b) (e) is for G-PCC, and (c) (f) is for AVS.}
  \label{Qsbar2MOSbar}
\end{figure*}

 The relationships between $\overline{MOS}$ and $Qs$ are illustrated in  Fig.~\ref{Qsbar2MOSbar}. For the attribute compression of all V-PCC, G-PCC and AVS compression algorithms, the relationship between $\overline{MOS}$ and $Qs$ follows the linear model, i.e.,
\begin{align}
\label{Eq5}
\overline{MOS}_a = c_{1,a}*{Qs}_a + c_{2,a},
\end{align}
where ${Qs}_a$ denotes the quantization steps for attribute compression. $c_{1,a}$ and $c_{2,a}$ are the model parameters, whose fitting values are shown in Table~\ref{tab:attrparm}.

\begin{table}[htbp]
\tiny
  \centering
  \caption{Fitting parameters in the attribute compression model.}
    \begin{tabular}{l|r|r|r}
    \hline
          & \multicolumn{1}{l|}{V-PCC} & \multicolumn{1}{l|}{G-PCC} & \multicolumn{1}{l}{AVS} \\
    \hline
    c1,a  & -0.0089 & -0.01 & -0.0519 \\
    c2,a  & 4.4862 & 5.3515 & 5.1337 \\
    \hline
    \end{tabular}%
  \label{tab:attrparm}%
\end{table}%

\subsection{Modeling the Geometry Compression}

\par For geometry compression of V-PCC, the relationship between $\overline{MOS}$ and ${Qs}$ follows the natural logarithm function, i.e.,
\begin{align}
\label{Eq6}
\overline{MOS}_{g,V-PCC} = c_{1,g}*ln{Qs}_g + c_{2,g},
\end{align}
where ${Qs}_g$ denotes the quantization steps for geometry compression. $c_{1,g}$ and $c_{2,g}$ denote the model parameters.

\par For geometry compression of G-PCC and AVS compression algorithms, the relationship between $\overline{MOS}$ and ${Qs}$ follows the linear model, i.e.,
\begin{align}
\overline{MOS}_{g,G-PCC,AVS} = c_{1,g}*{Qs}_g + c_{2,g}.
\end{align}
\par The fitted parameters in the geometry compression models are shown in Table~\ref{tab:geoparm}:

\begin{table}[htbp]
\tiny
  \centering
  \caption{Fitting parameters in the geometry compression model.}
    \begin{tabular}{l|r|r|r}
    \hline
          & \multicolumn{1}{l|}{V-PCC} & \multicolumn{1}{l|}{G-PCC} & \multicolumn{1}{l}{AVS} \\
    \hline
    c1,g  & -0.559 & -0.2381 & -0.273 \\
    c2,g  & 5.4165 & 5.3818 & 5.5034 \\
    \hline
    \end{tabular}%
  \label{tab:geoparm}%
\end{table}%

\subsection{Combining the Attribute Model and Geometry Model}

\par The point clouds are often compressed in both attribute and geometry, and the attribute degradation and geometry degradation are superimposed on the point clouds at the same time. As explored in Section \ref{combination}, the linear combination of the attribute model and geometry model can accurately estimate the quality. We take the weighted summation of ${MOS_a}$ and ${MOS_g}$ to predict the final quality scores.

\par For V-PCC, the established model is
\begin{align}
\label{EQ9}
{MOS_f} = p_{1,a}*{Qs_a} + p_{1,g}*lnQs_g + P.
\end{align}
\par For G-PCC and AVS, the established model is
\begin{align}
\label{EQ10}
{MOS_f} = p_{1,a}*{Qs_a} + p_{1,g}*Qs_g + P,
\end{align}
where $MOS_f$ is the predicted quality scores, $Qs_a$ is the quantization steps for attribute compression, and $Qs_g$ is the quantization steps for geometry compression. $p_{1,a} = \frac{1}{2} * c_{1,a}$, $p_{1,g} = \frac{1}{2} * c_{1,g}$, and $P = \frac{1}{2} * (c_{2,a} + c_{2,g})$ to cast the predicted quality scores under the same range of subjective scores.

\subsection{Analysis}

\par Some findings can be made in the experiment: i) Eq. \ref{Eq5} and Eq. \ref{Eq6} demonstrate that for the V-PCC distortion, the geometry distortion is more annoying compared with the attribute distortion, but the human eyes are more sensitive to the quantization change in the attribute compression; ii) for the geometry compression, the fitting curves of V-PCC and G-PCC are different, which derives from that the quantization of V-PCC is conducted on the projection while the quantization of G-PCC and AVS is conducted on octree; iii) for the attribute compression, all the three compression algorithms follow the linear model, since their quantization is all conducted on RGB, resulting in the similar perceptual pattern.

\par A potential concern is whether the obtained relation function is generic for different datasets. As discussed in Section \ref{overallmodel}, the difference of reference samples will only affect the additive factors, as $P$ in Eq. \ref{EQ9} and Eq. \ref{EQ10} which is a predefined constant. Thus, the obtained relation function can still accurately predict the quality rank of samples in other datasets, which is demonstrated by the cross-dataset evaluation in Section \ref{sec:crossexperiment}. 

\section{Experiments}
\label{sec:experiment}

\par In this section, we evaluate the performance of the proposed R-PCQA on the established PCCQA and WPC~\cite{Su2019WPC} dataset. Specifically, we use PCCQA to fit the model parameters and evaluate the fitting errors. Then, we evaluate on WPC dataset as cross check to verify the performance  of R-PCQA and its generalization ability.

\subsection{Error Analysis}

\par The proposed PCCQA consists of three parts, part 1: lossless-G-lossy-A, part 2: lossy-G-lossless-A and part 3: lossy-G-lossy-A. The proposed R-PCQA is fitted on the lossless-G-lossy-A and lossy-G-lossless-A parts, and we use the remaining lossy-G-lossy-A part to evaluate the performance. Especially, we note the former two parts as the training set and the latter part as the testing set. The mean, standard deviation and 95\% quantile of fitting errors $MOS - MOS_f$ on the testing set are shown in Table~\ref{tab:error}. The correlation performance on the testing set is shown in Table~\ref{tab:testing}.

\begin{table}[htbp]
\tiny
  \centering
  \caption{Mean, standard deviation and 95\% quantile of the fitting errors on the testing set.}
    \begin{tabular}{l|rrr}
    \hline
          & \multicolumn{1}{l}{Mean} & \multicolumn{1}{l}{Standard deviation} & \multicolumn{1}{l}{95\% quantile} \\
    \hline
    V-PCC & 0.0378 & 0.5885 & 0.7561 \\
    G-PCC & -0.5794 & 0.5113 & 0.0969 \\
    AVS   & -0.1356 & 0.2845 & 0.2531 \\
    \hline
    \end{tabular}%
  \label{tab:error}%
\end{table}%

\par We can see from Table~\ref{tab:error} and Table~\ref{tab:testing} that the proposed model can not only fit the dataset accurately, but also conforms to the characteristics of human visual system.

\subsection{Combination Analysis}
\label{combination}

\par The correlation performance of four combination schemes of the attribute model and geometry model on the testing set is shown in Table~\ref{tab:testing}.

\begin{table}[htbp]
\tiny
  \centering
  \caption{Correlation performance of four combination schemes on the testing set.}
    \begin{tabular}{l|rrr|rrr}
    \hline
          & \multicolumn{1}{l}{PLCC} & \multicolumn{1}{l}{SROCC} & \multicolumn{1}{l|}{RMSE} & \multicolumn{1}{l}{PLCC} & \multicolumn{1}{l}{SROCC} & \multicolumn{1}{l}{RMSE} \\
\cline{2-7}          & \multicolumn{3}{c|}{Linear Combination} & \multicolumn{3}{c}{Multiplicative Combination} \\
    \hline
    V-PCC & 0.8360 & 0.8554 & 0.5070 & 0.8360 & 0.8554 & 0.5070 \\
    G-PCC & 0.9854 & 0.9582 & 0.2098 & 0.9853 & 0.9582 & 0.2100 \\
    AVS   & 0.9917 & 0.9854 & 0.1650 & 0.9913 & 0.9854 & 0.1691 \\
    \hline
          & \multicolumn{3}{c|}{$G^{A}$ Combination} & \multicolumn{3}{c}{$A^{G}$ Combination} \\
    \hline
    V-PCC & 0.8351 & 0.8554 & 0.5082 & 0.8356 & 0.8554 & 0.5075 \\
    G-PCC & 0.9444 & 0.9582 & 0.4046 & 0.9767 & 0.9582 & 0.2644 \\
    AVS   & 0.9881 & 0.9854 & 0.1978 & 0.9862 & 0.9854 & 0.2127 \\
    \hline
    \end{tabular}%
  \label{tab:testing}%
\end{table}%

\par We can see from Table~\ref{tab:testing} that: i) the linear combination is determined due to its slightly better performance and simpler calculation for two relationship model mixing; ii) the combination schemes hardly affect the performance, which indicates that the obtained relationship models for attribute and geometry are independent. Due to the removal of mutual masking, it is not necessary to consider the interaction of attribute and geometry components in the mixing.

\subsection{Cross-dataset Evaluation}
\label{sec:crossexperiment}

\par After the model is established on the proposed dataset, we evaluate its generalization performance on another independent dataset, the V-PCC part of WPC~\cite{Su2019WPC}, which contains 400 distorted samples derived from 16 reference point clouds with 25 different quantization parameters. The results are shown in Table~\ref{tab:WPCVPCC}.

\begin{table}[htbp]
\tiny
  \centering
  \caption{Cross-dataset performance on WPC dataset.}
    \begin{tabular}{l|rr|l|rr}
    \hline
          & \multicolumn{1}{l}{PLCC} & \multicolumn{1}{l|}{SROCC} &       & \multicolumn{1}{l}{PLCC} & \multicolumn{1}{l}{SROCC} \\
    \hline
    M-p2po (FR)~\cite{cignoni1998metro} & 0.61  & 0.58  & H-PSNRyuv (FR)~\cite{MPEGSoft} & 0.29  & 0.23  \\
    M-p2pl (FR)~\cite{Mekuria2016Evaluation} & 0.63  & 0.59  & PCQM (FR)~\cite{meynet2020pcmd} & \textbf{0.74} & \textbf{0.75} \\
    H-p2po (FR)~\cite{cignoni1998metro} & 0.51  & 0.46  & GraphSIM (FR)~\cite{yang2020graphsim} & \textbf{0.74} & \textbf{0.75} \\
    H-p2pl (FR)~\cite{Mekuria2016Evaluation} & 0.55  & 0.48  & MPED (FR)~\cite{yang2021MPED} & 0.60  & 0.59  \\
    PSNRyuv (FR)~\cite{MPEGSoft} & 0.46  & 0.47  &       &       &  \\
    \hline
    PCM\_RR (RR)~\cite{Irene2020RR2} & 0.42  & 0.38  & \textbf{R-PCQA (RR)} & \textbf{0.88} & \textbf{0.88} \\
    \hline
    \end{tabular}%
  \label{tab:WPCVPCC}%
\end{table}%

\par We can see from Table~\ref{tab:WPCVPCC} that: i) for the compression distortions, the proposed R-PCQA exhibits the state-of-the-art performance which only needs the assistance of two compression parameters, even compared with the existing FR metrics; ii) the model parameters derived from PCCQA still exhibit robust performance on another independent dataset, which demonstrates the generalization ability of the proposed RR metric R-PCQA; iii) the massive increase in points after reconstruction may interfere with the measurement of point-wise FR metrics, resulting in the poor performance of some FR metrics.

\section{Conclusion}
\label{sec:conclusion}

\par In this paper, we analyze the compression distortions of point clouds under separate attribute compression and geometry compression to avoid their mutual masking. Then by fitting the relationship between the quantization steps and the perceptual quality, we propose a RR PCQA model, called R-PCQA, for evaluating V-PCC, G-PCC and AVS distortions. The experiment results have demonstrated that the proposed R-PCQA exhibits reliable and robust performance.

\section{Acknowledgement}

This paper is supported in part by National Key R\&D Program of China (2018YFE0206700), National Natural Science Foundation of China (61971282, U20A20185). The corresponding author is Yiling Xu (e-mail: yl.xu@sjtu.edu.cn).



%

\bibliographystyle{IEEEtran}
\bibliography{refs,ref_ICIP}

\end{document}